%% file: 00-main.tex
\documentclass[conference]{IEEEtran}

\input{01-packages}

\begin{document}

\input{02-top-matter} 

\maketitle

\input{03-abstract.tex}



\input{10-intro}

\input{20-algorithm}
\input{30-results.tex}
\input{40-conclusion}

\input{90-acks} 
\small
\bibliographystyle{IEEEtran}
\bibliography{refs}

\end{document}

%% file: 01-packages.tex
\usepackage[bookmarks=false]{hyperref}
\usepackage[dvipsnames]{xcolor} 
\usepackage{wrapfig}
\usepackage{booktabs}

\usepackage{amsmath, amssymb, amsfonts, amscd} 
\usepackage{amsthm}
\usepackage[justification=raggedright]{subcaption}
\usepackage{courier} 
\usepackage{graphicx}
\graphicspath{ {images/} } 
\usepackage{multicol}
\usepackage{lipsum} 
\usepackage[numbers,sort&compress,square]{natbib} 
\usepackage{enumitem} 
\usepackage{todonotes}

\usepackage{listings}
\usepackage{inconsolata}

\usepackage{threeparttable}
\usepackage{array}
\usepackage{multirow}

\newcommand\norm[1]{\left\lVert#1\right\rVert}

\usepackage[ruled]{algorithm}
\usepackage[noend]{algpseudocode}
\makeatletter
\renewcommand{\ALG@beginalgorithmic}{\small}
\makeatother
\usepackage{tabularx}

\IEEEoverridecommandlockouts 

\usepackage{outlines}
\usepackage{dblfloatfix}

%% file: 02-top-matter.tex
\title{\LARGE{ACTT: Automotive CAN Tokenization and Translation}
\thanks{\footnotesize{This manuscript has been authored by UT-Battelle, LLC under Contract No. DE-AC05-00OR22725 with the U.S. Department of Energy. The United States Government retains and the publisher, by accepting the article for publication, acknowledges that the United States Government retains a non-exclusive, paid-up, irrevocable, world-wide license to publish or reproduce the published form of this manuscript, or allow others to do so, for United States Government purposes. The Department of Energy will provide public access to these results of federally sponsored research in accordance with the DOE Public Access Plan (\url{http://energy.gov/downloads/doe-public-access-plan}).} }
}

\author{
\IEEEauthorblockN{Miki E. Verma, Robert A. Bridges, Samuel C. Hollifield}\\
\IEEEauthorblockA{Cyber \& Applied Data Analytics Division, Oak Ridge National Laboratory, Oak Ridge, TN\\
\{vermake, bridgesra, hollifieldsc\}@ornl.gov}}

%% file: 03-abstract.tex
\begin{abstract}
Modern vehicles contain scores of Electrical Control Units (ECUs) that broadcast messages over a Controller Area Network (CAN).
Vehicle manufacturers rely on security through obscurity by concealing their unique mapping of CAN messages to vehicle functions which differs for each make, model, year, and even trim. 
This poses a major obstacle for after-market modifications notably performance tuning and in-vehicle network security measures.
We present ACTT: Automotive CAN Tokenization and Translation, a novel, vehicle-agnostic, algorithm that leverages available diagnostic information to
parse CAN data into meaningful messages, 
simultaneously cutting binary messages into tokens, and learning the translation to map these contiguous bits to the value of the vehicle function communicated.

\end{abstract}

%% file: 10-intro.tex
\section{Introduction \& Background}
\label{sec:intro}
Modern vehicles rely on dozens to greater than a hundred electronic control units (ECUs), embedded computers that send periodic messages to orchestrate sub-component functionality, including life-critical services, such as brakes. 
ECUs broadcast messages over a Controller Area Network (CAN), which defines a lightweight protocol that is efficient and functionally sound, but lacks security measures such as authentication and encryption \cite{Koscher_Experimental_2010}.  
After-market efforts involving modern vehicles, e.g., performance tuning or  adding security measures, usually require the ability to interact with and understand the data sent over this in-vehicle network. 

\begin{figure}[t]
\vspace{-.2cm}
\includegraphics[width=.48\textwidth]{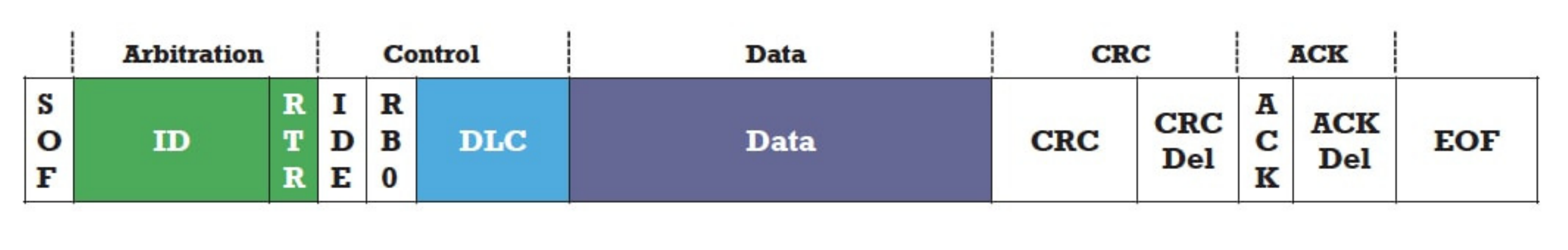}
\caption{
Image from Cho \& Shin \cite{cho2016fingerprinting} of CAN 2.0 data frame. Arbitration ID used for indexing and prioritizing the packet, which contains a 64 bit data field or message.
}
\label{fig:CANframe}
\end{figure}

The CAN 2.0 specification defines aspects of the physical and data link layer, particularly the CAN frame format, which is standard across all implementations, and is publicly available \cite{Bosch_GmbH_1991}. 
CAN frames (depicted in Figure \ref{fig:CANframe}) have several components
but there are only two important components to understand the frame\textemdash the arbitration ID (AID), which is used to identify the packet as well as determine priority, and the data field, containing up to 64-bits of message contents. 

However, unlike the open specification of the physical and data link layer, the full specification of how to decode the data field is completely proprietary; it is held secret by the original equipment manufacturer (OEM) and
varies per make, model, year, and trim.  
This proprietary information is generally stored in a \texttt{.dbc} file format (database for CAN).
DBCs contain (1) signal definitions, which describe the following: 
the segment position (start and end bit indices);
the binary-to-decimal encoding scheme (signed vs. unsigned, little vs. big endian);
and the conversion information for translating the decimal into a meaningful physical value (offset and scale factor, units, and range of possible values). 
The DBC also includes (2) message timing attributes with information such as transmission frequency (how often message with particular AID is sent), whether this rate is constant or triggered by an event, etc., and (3) the ECU that sent the message. 
Nearly all after-market modifications or research on passenger vehicles requires and critically relies on reverse engineering some of the information held in the DBC. 
For example, much of current CAN intrusion detection research is based on data-driven approaches to determine  (2) AID message timing (e.g. \cite{moore2017modeling, gmiden2016intrusion}) or (3) ECU identification (e.g. \cite{cho2016fingerprinting, choi2018identifying}), by leveraging physically observable characteristics, namely message timestamps and voltage (physical layer).


\subsection{Problem Statement} 

\begin{wrapfigure}[11]{r}{.25\textwidth}
\vspace{-.5cm}
    \centering
    \includegraphics[width=.25\textwidth]{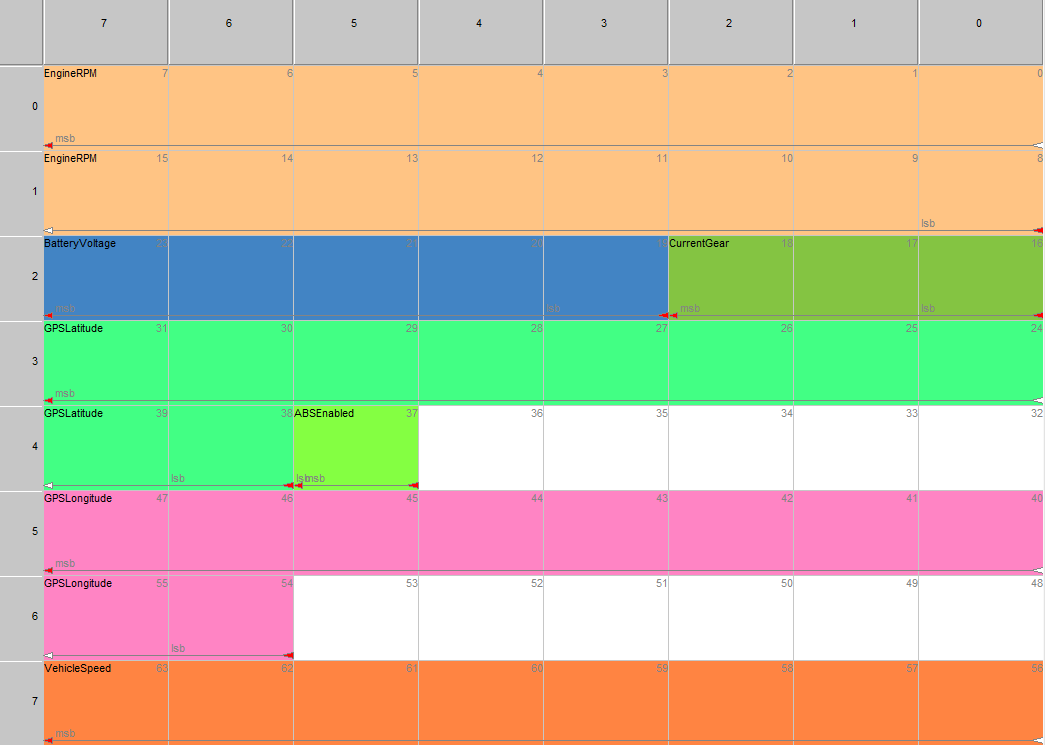}
    \caption{64-bit CAN Message Description defined in DBC: signals depicted in color with unused bits depicted in white. Produced by DBC editor \textit{CANdb++} (\url{www.vector.com/candblib}).}
    \label{fig:dbc_eg_layout}
\end{wrapfigure}
However, reverse engineering (1) the signal definitions 
is a significantly more difficult problem than (2) and (3). One can monitor and send CAN packets,
but understanding how to translate the data field 
information to what it encodes is not currently possible. 
An example of a fully defined message with signals shown in color is depicted in Figure \ref{fig:dbc_eg_layout}, which was produced by software for editing DBCs. 

The problem of defining a group of signals held in a 64 bit data field is two-fold: 
\begin{enumerate}[leftmargin = 0.5cm] 
\item {\it Tokenization:} segmenting message contents into {\it tokens}, contiguous sequences of bits, that constitute a single signal (determining start/end bit, binary encoding scheme)
\item {\it Translation:} converting the token into a number and understanding its meaning in terms of the vehicle's function, e.g. front left wheel speed, brake light on, etc. (determining offset, scale factor, units, value range)
\end{enumerate}
Thus, a \textit{signal} is a \textit{token} paired with a meaningful mapping, or \textit{translation}.




\subsection{Vehicle Information \& Diagnostics}
While tokenizing and translating CAN signals is not readily available for passenger vehicles, other ground-truth vehicle information has been leveraged. 
Some research opts for hand-labeled drive states (i.e accelerate, reverse, key in, speedometer reading of 20mph, ect.) \cite{tyree2018exploiting, Jaynes_Dantu_Varriale_Evans_2016, Narayanan_Mittal_Joshi_2015}, while others used external data loggers 
\cite{Huybrechts_Vanommeslaeghe_Blontrock_Van_Barel_Hellinckx_2018}. 


One particularly good source of data is automotive diagnostic data.  
Vehicles manufactured in 2008 or later have an on-board diagnostic (OBD-II) port, allowing for open access to automotive networks, and mandated 
for state-wide emission testing and diagnostics by the J1979 standard.
Automotive diagnostics exist separately as an application layer for CAN implementations\textemdash for diagnostic actions, the application layer is called the Unified Diagnostic Service (UDS). 
UDS exists as a request-response system in which ECUs respond to interrogation regarding a variety of vehicular states. One can query for data such as engine speed, wheel speed, oxygen sensor readings, each corresponding to a particular Diagnostic OBD-II PID (DID), for which units, ranges, and conversion formulas are public \cite{obd2pids}.
Importantly, these CAN signals exist in addition to the normal CAN traffic which the vehicle uses for critical functions, although both are seen in the same data stream.
Further, this serves as a reliable way to obtain ground-truth data without the need for exogenous data streams.

\subsection{Related Works}
\label{sec:related-works} 
Recent research is emerging to provide solutions for signal extraction from passenger vehicle CAN data fields.  
Related works either attempt unsupervised tokenization of CAN data frames or leveraging diagnostic data, but not directly for extracting CAN signals.

Markowitz \& Wool \cite{Markovitz_Wool_2017} work toward anomaly detection on CAN data, including a method to tokenize CAN data and provide high-level categorization of these tokens (akin to part-of-speech tagging tokenized text data). 
Focusing on reverse engineering automotive CAN data, Marchetti \& Stabili~\cite{Marchetti_Stabili_2018} refine the algorithm of Markovitz and Wool \cite{Markovitz_Wool_2017}, define the semantic categories more rigorously, and test the the method results against a DBC they acquired. 
Neither works give explicit mappings of CAN message segments to their meanings, but do provide an unsupervised method for tokenization and semantic categorization.
Concurrent work by Nolan et. al.~\cite{Nolan_Mullins_Graham_Kabban} develops a similar method for unsupervised tokenization of CAN data frames, based on bit flip probabilities. 
The elegant method simply partitions the 64-bit message frames into appropriately sized segments. 





Similar to our approach, previous works have focused on leveraging UDS data as ground truth for analysis, e.g.,~\cite{Flach_Mishra_Pedrosa_Riesz_Govindan_2011, Wasicek_Pese_Weimerskirch_2017, Li_Zhao_Juliato_Ahmed_Sastry_Yang_2017}, although they did not address the explicit problem of CAN data  interpretation. 
Li et al.~\cite{Li_Zhao_Juliato_Ahmed_Sastry_Yang_2017} presented an IDS that used a regression model to learn relationships between physical values, such as vehicle speed, and raw CAN data, whereas Wasicek et al. ~\cite{Wasicek_Pese_Weimerskirch_2017} develop an IDS based solely on anomalies in diagnostic data correlations. 
Neither address the problem of actual semantic analysis, and both would require UDS commands to be fired continually during training and IDS deployment. 

Huybrechts et al.~\cite{Huybrechts_Vanommeslaeghe_Blontrock_Van_Barel_Hellinckx_2018} is the only previous work that applied UDS annotations towards CAN data translation.
They developed an ``arithmetic method'' that attempts to label sections of data fields based on similarity to simultaneously collected diagnostic data. 
However, they did not  address tokenization, instead  considering segmentation by one- or two-byte tokens (a shortcoming of their method that they acknowledge), and they provide no explicit linear mapping to translate segments to real physical values.

\subsection{Contributions} 
\label{sec:contributions} 

We provide a workflow for collecting CAN data alongside available diagnostic information and a novel algorithm,  ACTT: Automotive CAN Tokenization and Translation, the first algorithm to bring together these previously separate streams of research. 
The contributions are as follows: 
\begin{itemize}[leftmargin = .3cm] 
\item ACTT uses diganostic labels to learn CAN signal definitions, including the parameters needed for tokenization (start/end bit, endianess) and translation (offset, scale factor, units, value range).
\item ACTT furnishes goodness-of-fit scores allowing visibility into what diagnostic codes are directly encoded in CAN data, and allowing discovery of CAN signals that are related but not directly accessible via diagnostic codes (e.g., accelerator depressed indicator). 
Furthermore, the scoring permits quantifying the percent of the CAN data field translated.   
\item ACTT provides a tuneable parameter that on one extreme forces extraction of only near-perfect diagnostic signals while on the other provides less exact matches clustered by their correlations. 
\end{itemize}


\subsection{Impact}
Our work also will aid other streams of related research, providing a preprocessing step and allowing them to refine and more rigorously test their methods.
Due to the difficulty of extracting CAN signals,
many current vehicle research efforts 
have avoided fully and explicitly determining these signals.  One workaround for this was to manually reverse engineer a few single-function signals in CAN data, e.g., \cite{Jaynes_Dantu_Varriale_Evans_2016}, in order to provide proof of concept. Others implemented machine learning methods (Hidden Markov Models, Neural Nets, Manifold Learning, etc.) that implicitly learn relationships between raw binary data and vehicular states \cite{moore2018data, tyree2018exploiting, Jaynes_Dantu_Varriale_Evans_2016, Narayanan_Mittal_Joshi_2015,Huybrechts_Vanommeslaeghe_Blontrock_Van_Barel_Hellinckx_2018,Wasicek_Pese_Weimerskirch_2017, Li_Zhao_Juliato_Ahmed_Sastry_Yang_2017}.

The features used in these machine learning methods were often based on unprincipled decisions regarding tokenization, e.g.,  considering each byte pair in the data field to be a ``signal''. Additionally, supervised methods required vehicle states that were often hand labeled, or relied on other exogenous data sets.
Our work therefore provides a preprocessing step for these methods that previously employed these brute force reverse engineering techniques, brittle tokenization schemes, or manual labeling of vehicle states. 

For after market tools, knowledge of signal definitions has been shown to be very beneficial in various streams of research. 
Heavy-duty vehicles' CANs follow the J1939 standard \cite{J1939_201206}, which is not proprietary and is like a standardized DBC for all trucks.
This largely facilitates rapid development of new features which can be vehicle agnostic. An 
example of cross-platform integration made possible by the J1939 standard is the Bendix Wingman Advanced which brings adaptive cruise control with braking features along with collision mitigation technology to a variety of trucks\footnote{\url{www.bendix.com/en/products/acb/wingmanadvanced_1.jsp}}.
For passenger vehicles, Ford has developed an open source API called OpenXC, which includes a small sample of signal definitions. 
This has resulted in a wealth of research by individuals creating add on tools such as `OpenXCThenThat', a vehicular task automator, and `Smart Battery App', an EV battery optimizer based on terrain, both created during the Ford Electrified Vehicle Hackathon, an event meant to show off the potential of OpenXC\footnote{\url{http://openxcplatform.com/}}. 

Overall, the  inability to comprehend passenger vehicle CAN data fields severely limits the range and effectiveness of after-market vehicle research and engineering. 
Finding vehicle-agnostic methods for syntactic and semantic understanding of CAN data fields promises a wealth of opportunity for after-market development.

%% file: 20-algorithm.tex
\smallskip
\begin{algorithm}[b]
  \caption{Token Preprocessing}\label{preprocess_tokens}
  \label{alg:1} 
  \begin{algorithmic}
    \Function{CategorizeBits}{$X$}
    \State $B_0, B_1, B_{used} \gets \emptyset$
    \For{$j\gets 0, \dots, 63$}
    \If{$X_{i,j} = 0$ $\forall i = 1,\dots,n $}
        \State $B_{0} \gets B_{0} \cup \{j\}$
    \ElsIf{$X_{ij} = 1$ $\forall i = 1,\dots,n$}
        \State $B_{1} \gets B_{1} \cup \{j\}$
    \Else
        \State $B_{used} \gets B_{used} \cup \{j\}$
    \EndIf
    \EndFor
    \State \textbf{return} $B_0, B_1, B_{\text{used}}$
    \EndFunction
  \end{algorithmic}
  \begin{algorithmic}
  
    \smallskip
  
  \Function{GetValidTokenBoundries}{$B_0, B_1, B_{\text{used}}$}
    \State $V \gets \{[j_s, j_e] \mid j_s, j_e \in B_{\text{used}} \text{ and }  j_s \leq j_e \text{ and }  [j_s, j_e] \cap B_{0} \cap B_{1} = \emptyset\}$
    \State \textbf{return} $V$
    \EndFunction
  \end{algorithmic}
\end{algorithm}

\section{Algorithm}
\label{sec:algorithm}
We assume we have a CAN capture from a vehicle during a sufficiently long driving period to exercise most variation in the CAN data. 
Here we define the notation for representing the 64-bit payloads for an AID over time, and then our algorithm for tokenization and translation.  


Let $X \in \left\{0,1\right\}^{n \times 64}$ be an \textit{AID trace}, a sequence of $n$  time-ordered 64-bit messages from the same AID, where $X_{ij}$ denotes the bit $j$ in the $i$\textsuperscript{th} message of the sequence, and messages occur at times $[t_1,\dots,t_n]$.

Let $y \in \mathbb{Z}^m$ be a \textit{Diagnostic trace}, a sequence of $m$ integer responses from the same DID, where messages occur at times  $[s_1,\dots,s_m]$. 
Note that unlike for AID traces, whose lengths can differ significantly based on priority and transmission rates, all diagnostic traces should be about length $m$. 
We note that we do not include constant diagnostic responses, that is DID traces s.t. $y(s_i)=c$  $\forall i=0,...,m$ are not considered. 

For $i = 1, \dots, n$, and $1\leq j_s \leq j_e \leq 64$, we define the little-endian (ending bit, $j_e$, is most significant bit) and big-endian (starting bit, $j_s$, is most significant bit) integer encodings of the bit subsequence $[X_{i,j_s}, \dots, X_{i,j_e}]$ as, respectively,
\begin{align}
\label{eq:lendien} 
    L(i, j_s, j_e)&= \sum_{j = j_s}^{j_e} X_{i,j} 2^{j-j_s}\\
\label{eq:bendien}
   B(i, j_s, j_e)&= \sum_{j = j_s}^{j_e} X_{i,j} 2^{j_e-j}\ .
\end{align}



Our algorithm consists of three main components: (1) Token Preprocessing, (2) Diagnostic Matching, (3) Message Packing,  discussed below and described in Algorithms \ref{alg:1}, \ref{alg:2}, and \ref{alg:3}, respectively.

\subsection{Token Preprocessing}
\label{sec:token-preprocess} 

We first examine the bits in each AID trace $X$ and categorize each into: a constant 1s bit, a constant 0s bit, or a `used' bit. 
We note that it is impossible to differentiate between a bit that is defined to be held constant as a buffer between signals (an unused bit) and a bit that is simply unchanged  during data collection due a state not being reached in the CAN capture under investigation. 
Letting $B_0, B_1, B_{\text{used}}$ denote the sets of bit positions, we can then determine the set of possible valid token boundaries $V$ (see Algorithm \ref{alg:1}). 
We consider valid tokens to be any contiguous set of bits that does not include a constant bit. Note that we have defined start and end bit indices $j_s, j_e$ to be inclusive (i.e. $j_s = j_e$ indicates a 1-bit token). 

\subsection{Diagnostic Matching}
\label{sec:diag-match-alg} 
We next determine whether any valid token of an AID’s 64-bit data field is related to a diagnostic response message in Algorithm \ref{alg:2} by converting a time-varying sequence of bit strings to a sequence of integers, then regressing to see if they linearly fit any time-varying diagnostic sequence collected. 

The first step is to determine the integer translation for each valid tokens string (from Algorithm \ref{alg:1} which returns set $V$ of possible start and end bits, $[j_s, j_e]$ of a non-constant token). 
We consider both little- and big-endian unsigned encodings, but do not consider any alternative encodings at this time (e.g. signed binary, one's complement, two's complement). 
Specifically, for each time index $i$, we convert $[X_{i, j_s}, \dots, X_{i, j_e}]$ (a sequence with each element a vector of bits) to two sequences of integers using Equations \ref{eq:lendien} \& \ref{eq:bendien}. 

For each of these endian encoding of the token trace, say $\mathbf{x} = [x_1, ..., x_n]$,  we use linear regression to find constants transforming $\mathbf{x}$ to each diagnostic trace $\mathbf{y} = [y(s_1), ..., y(s_m)]$; 
of course, we first interpolate the token trace $\mathbf{x} = [x_1, \dots, x_n]$ to the $m$ diagnostic time points $[s_0,\dots,s_m]$ giving $\mathbf{\tilde{x}} = [\tilde{x}_1, \dots, \tilde{x}_m]$.  
The regression furnishes the coefficients $a, b$  that results in the best linear fit  $\mathbf{\hat{y}}_{\{a,b\}} = a \mathbf{\tilde{x}} + b$ to $\mathbf{y}$. 
Note that we choose to interpolate over the $m$ diagnostic points because this is sampled at a much lower rate than AID messages occurring in normal CAN traffic. 
We then score each model using the coefficient of determination, $R^2$ (see Algorithm \ref{alg:2} for formula), and add the token to our match set $M$ if the score exceeds a set threshold $\alpha$. 
Recall $R^2\in (-\infty, 1]$ with $R^2 = 1$ indicating perfect fit, and $R^2 = 0$ indicating fit of a horizontal line.

\begin{algorithm}[t]
  \caption{Diagnostic Matching}
  \label{alg:2}
  \begin{algorithmic}
  \Function{MakeIntegers}{$X, (j_s, j_e), Endianness$}
        \For{$i \gets 1, ..., n$}
            \If{$Endianness = $ `little-endian'}
                \State $x_{i} \gets \sum_{j=j_s}^{j_e}X_{i,j}2^{j-j_s}$
            \Else
                \State $x_{i} \gets \sum_{j=j_s}^{j_e}X_{i,j}2^{j_e-j}$ 
            \EndIf
        \EndFor 
    \State $\mathbf{x} \gets [x_1, \dots, x_n]$
    \State \textbf{return} $\mathbf{x} $
  \EndFunction
  \end{algorithmic}
  
  \smallskip
  
  \begin{algorithmic}
  \Function{CoefDeterm}{$\mathbf{y}, \mathbf{
  \hat{y}}$}
    \State $S_{tot} \gets \sum (y_i - \text{ave}(\mathbf{y}))$
    \State $S_{res}  \gets \sum (\hat{y}_i - y_i)$
    \State $R^2 \gets 1 - (S_{res} / S_{tot})$  
    \State \textbf{return} $R^2$
  \EndFunction
        
  \end{algorithmic}

  \smallskip 
  
  \begin{algorithmic}
  \Function{LinearFit}{$\mathbf{x}, \mathbf{y}$}
    \State $\mathbf{\tilde{x}} \gets $\Call{Interpolate}{$\mathbf{x}, \mathbf{y}$} 
        \State  ( Denote $\mathbf{x} = [x_1, \dots, x_n]$, $\mathbf{y} = [y(s_1), \dots, y(s_m)]$ and
        \State interpolate over diagnostic times:   $\mathbf{\tilde{x}}=[x(s_1),\dots x(s_m)]$ )
    \State Find best fit $\mathbf{\hat{y}}_{\{a,b\}}$: $\min_{a,b} \norm{a\mathbf{\tilde{x}} + b - \mathbf{y}}_2^2$
    \State $R^2 \gets $\Call{CoefDeterm}{$ \mathbf{y}, \mathbf{\hat{y}}_{\{a,b\}}$}
    \State \textbf{return} $R^2, a, b$
  \EndFunction
        
  \end{algorithmic}
  
    \smallskip
  
  \begin{algorithmic}
  \Function{MatchTraces}{$X, V, \mathbf{y}, \alpha$}
    \For{ $(j_s, j_e) \in V$ }
        \For{ $Endianness \gets $ [`little-endian', `big-endian']}
            \State $\mathbf{x} \gets$ \Call{MakeIntegers}{$X, (j_s, j_e), Endianness$}
            \State $ R^2, a, b \gets $ \Call{LinearFit}{$\mathbf{x}, \mathbf{y}$}
            \If{$R^2 \geq \alpha$} 
                \State $M \gets M \cup \{(j_s,j_e, Endianness, R^2, a, b)\}$ 
            \EndIf
        \EndFor
    \EndFor
    \State \textbf{return} $M$
  \EndFunction
  \end{algorithmic}
  
\end{algorithm}

There are several important things to note about the matching algorithm. 
Firstly, the goodness of fit for each model gives more than just a score of how well the model performs, but an indication of whether the token actually encodes the diagnostic signal. 
In some cases it indicates that the CAN token is not an exact match, but does  encode a correlated signal, e.g., we have identified 1-byte tokens with high correlation to a continuously changing DID, presumably a binary indicator. 
Hence $\alpha$ can be tightened to isolate near-perfect encodings, or tuned down to discover multiple related signals to each DID. 

Secondly, we get both the tokenization (AID, start and end bits $[j_s, j_e]$) and the actual mapping ($a, b$, endian-ness) needed to translate the binary message to the decimal value. Note that in order to get the true value measuring a physical signal, including units, we must apply the conversion for the matched diagnostic response, a formula accessible for these standardized DIDs \cite{obd2pids}.
We note that changing $\alpha$ can change the token boundaries, due to the fact that the algorithm is simultaneously learning both tokenization and translation.

Thirdly, the scoring mechanism is quite flexible.
It can be easily altered to consider other binary encodings, more flexible regression, or other time-varying observations in addition to the DIDs. 



\subsection{Message Packing}
\label{sec:packing} 
We now have candidate tokens and their mappings, but it is possible for these learned tokens to overlap, (e.g., the token comprised of bits 2 to 8 map to a DID with a high fit and the token comprised  of bits 1 to 4 map to a different DID with a high fit),
leaving a problem of deciding which to choose for each AID. 
The final section of the algorithm approaches this problem by determining the optimal packing of matched tokens for each AID data field. 
The goal is to essentially create as full a DBC data field description as seen in Figure \ref{fig:dbc_eg_layout} as possible, choosing the tokens with high goodness of fit to a DID and preferring longer tokens to shorter. 
For a given match set from $M$, we scale the coefficient of determination, $R^2$, by the token length $j_e-j_s +1$ and then optimize the sum of these scores over non-overlapping matched tokens (see Algorithm \ref{alg:3}).  
While we do not provide the full algorithm, we note that this does not require an exponential-time algorithm\textemdash the globally optimal solution can be found in $\mathcal{O}(n\log{n})$ time using dynamic programming (see \cite{weighted_interval_sched} for full weighted interval scheduling dynamic programming algorithm). 

\begin{algorithm}[t]
  \caption{Message Packing Algorith/Score: Modeling token packing as a weighted interval scheduling problem, this can be solved using dynamic programming in $\mathcal{O}(n\log{n})$ assuming first sorting by end-bit index $j_e$.}
  \label{alg:3} 
  \begin{algorithmic}
  \Function{FindOptimalPayload}{$M$}
    \State Following \cite{weighted_interval_sched} find and return $ \max$ and $\text{argmax}$ of\\
    $(1/64) \sum_T R^2 (j_e-j_s+1) : T = \{(j_s, j_e, R^2, \cdot) \in M \mid \bigcap[j_s, j_e) = \emptyset \}$  
  \EndFunction
  \end{algorithmic}
\end{algorithm}
We refer to the maximum of the sum in Algorithm \ref{alg:3} as the \textit{message packing score} taking values in $[0,1]$. 
Note that if all 64 bits are used by tokens matched with perfect fit ($R^2 = 1$), then the score is 1. 
Observing this score over the percent of non-constant or `used' bits gives a measure of how successful the tokenization and translation process is. 
Examples provided in Results Section. 

%% file: 30-results.tex
\section{Results}
\label{sec:results} 
We tested our method on three vehicles dated 2008, 2015, 2016 of two different makes, three models, and using both gasoline and hybrid.
For this short paper we present results and examples from a 20 minute capture from a 2008 gasoline vehicle in city and highway driving conditions. 
CAN traffic was captured using a Kvaser Leaf Lite V2 (\url{www.kvaser.com}) providing CAN-to-USB translation to a Linux OS laptop using SocketCan software (\url{https://elinux.org/Can-utils}). 
This particular car used 25 AIDs and responded on 31 DIDs, which were queried at a rate of 20Hz throughout the capture.
OEMS are only required to have their vehicles respond to a subset of about 200 possible DIDs,
and for the cars we tested, we found similar subsets of about 30-45 responsive DIDs. 
We note that we obtained similar, if less comprehensive results for all vehicles tested (2015, 2016 cars had about two times as many AIDs) and that results were similar across multiple captures. 
We ran the tokenization and translation algorithm with $\alpha = 0.50$.
 
\begin{figure}[]
\vspace{-.4cm}
\caption{Examples of mapped messages for three AIDs. Maps similar to DBC layout in Figure 
\ref{fig:dbc_eg_layout} shown with colored adjacent bits showing matched tokens annotated with matched DID. Constant 0s and 1s bits are shown in grey and black respectively. Plots of each matched and unmatched token are also plotted for the first 5 min of each trace.
For matched tokens, AID token values $\mathbf{\hat{y}}_{\{a,b\}}$ are plotted at diagnostic time points in matched color on top of the diagnostic curve $\mathbf{y}$. Score, coefficients, $a,b$, endianness (see Algorithm \ref{alg:2}) is also given. Unmatched AID tokens are plotted in black at AID time points.}
\label{fig:mappedpayloads}

\vspace{.2cm}
\begin{subfigure}[c]{0.48\textwidth}
    \centering
    \includegraphics[width =\textwidth]{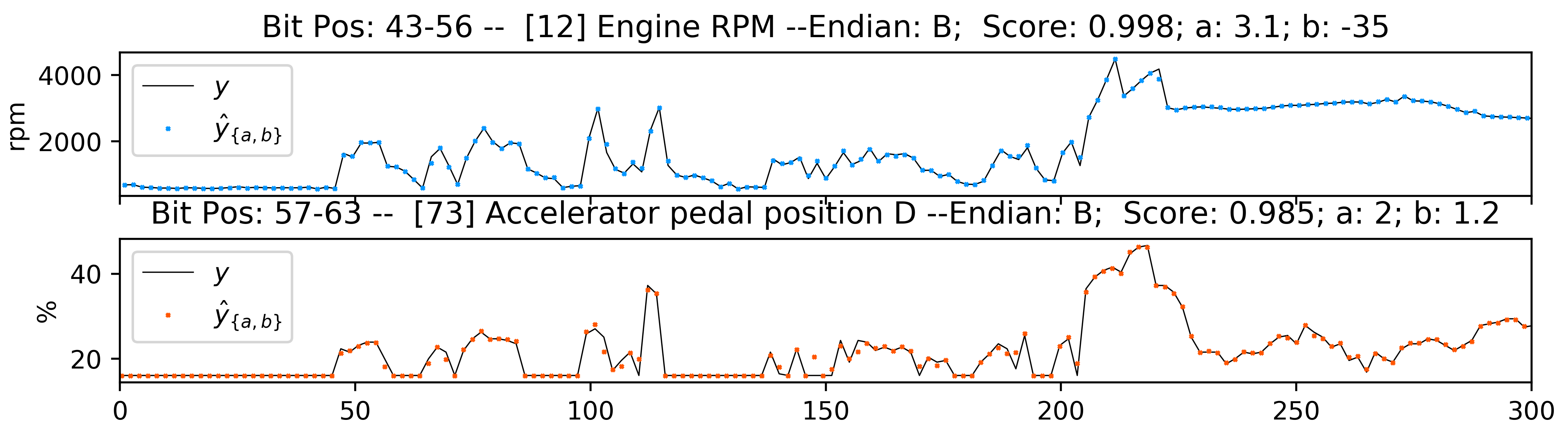}
  \begin{minipage}[c]{0.36\textwidth}
    \includegraphics[width =\textwidth]{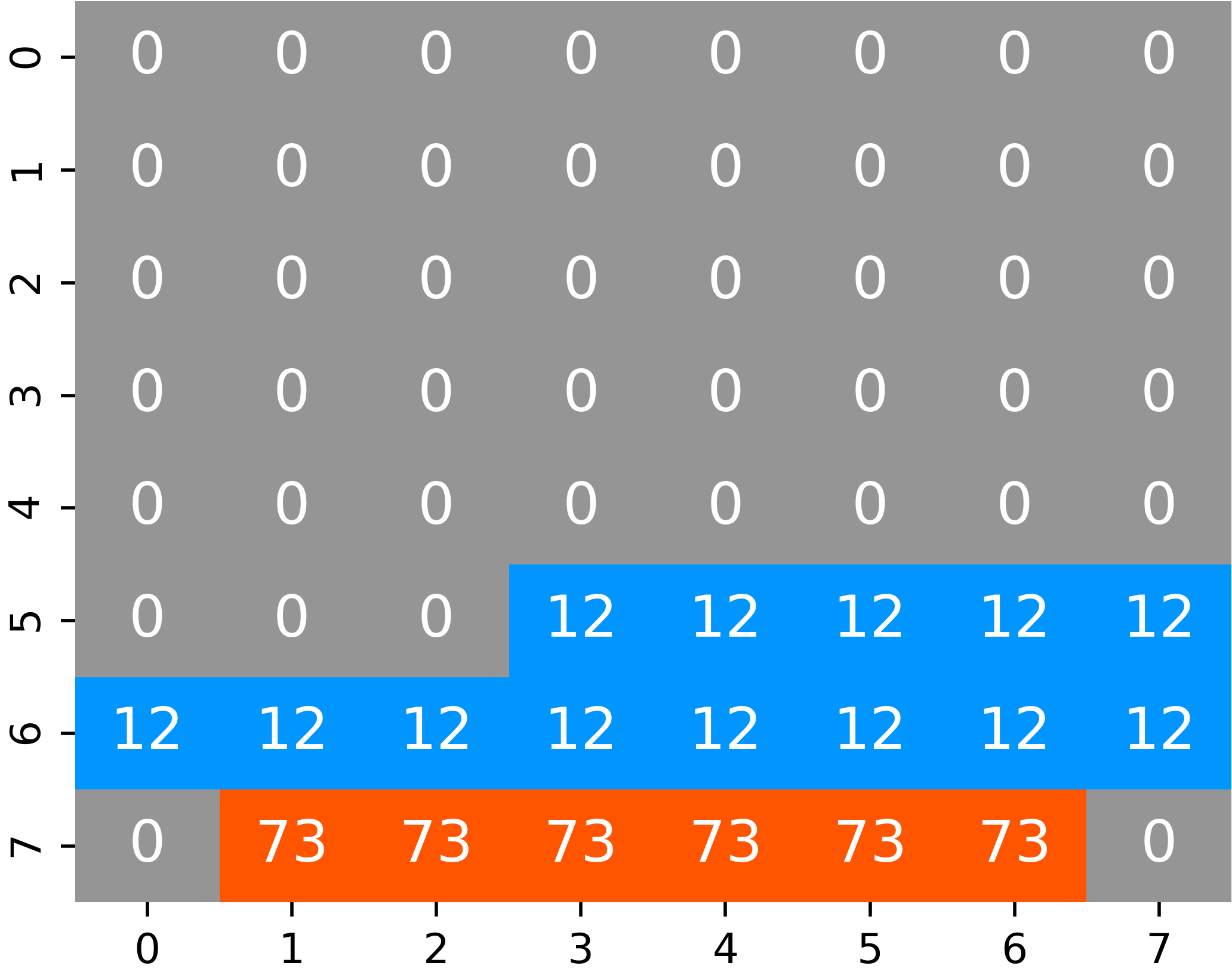}
  \end{minipage}
  \begin{minipage}[c]{0.62\textwidth}
  \caption{AID18: Fully mapped lower priority message encoding Engine RPM (DID12) and Accelerator Pedal Pos. (DID73) in the last three bytes. The top three bits in byte five are unused, likely because we did not go fast enough to get near maximum RPM. The high match scores $(>.98)$ reflect the almost perfect fit seen by visual inspection.}
  \label{fig:mp-top} 
  
\end{minipage}
\end{subfigure}

 \hrule
\vspace{.2cm}

\begin{subfigure}[c]{0.48\textwidth}
    \centering
    \includegraphics[width =\textwidth]{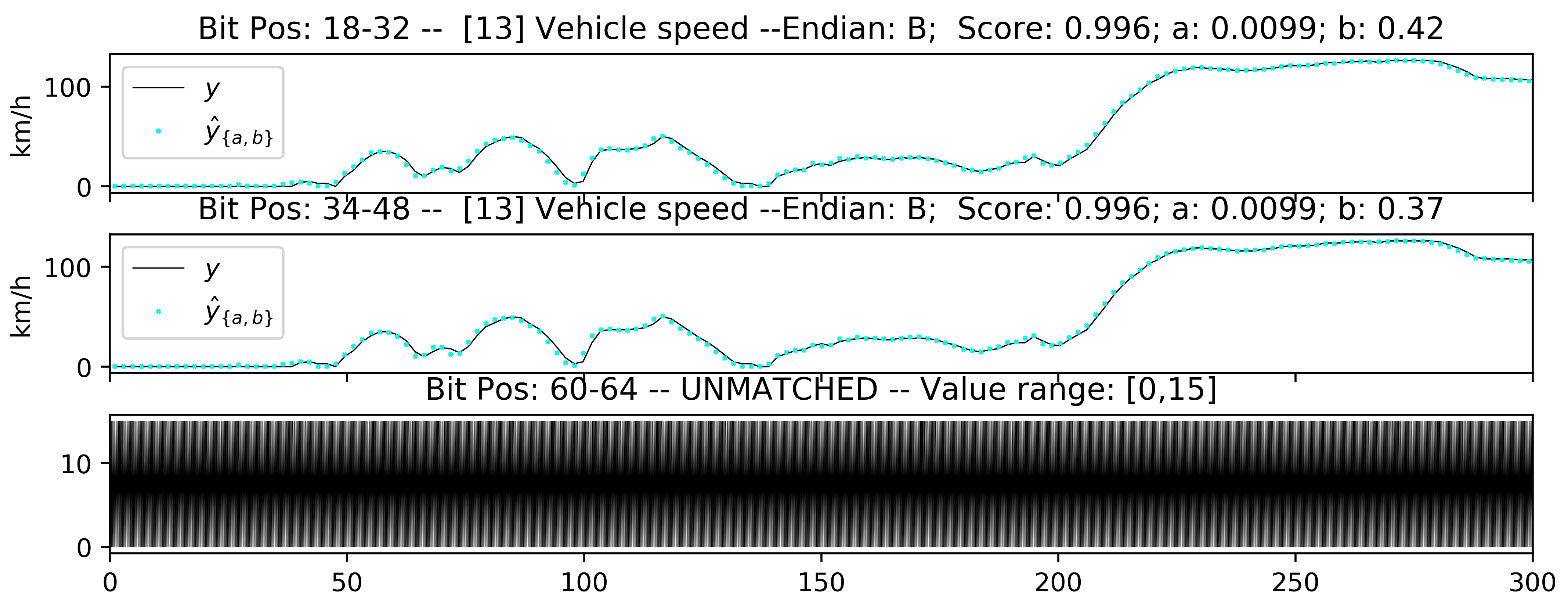}
  \begin{minipage}[c]{0.36\textwidth}
    \includegraphics[width =\textwidth]{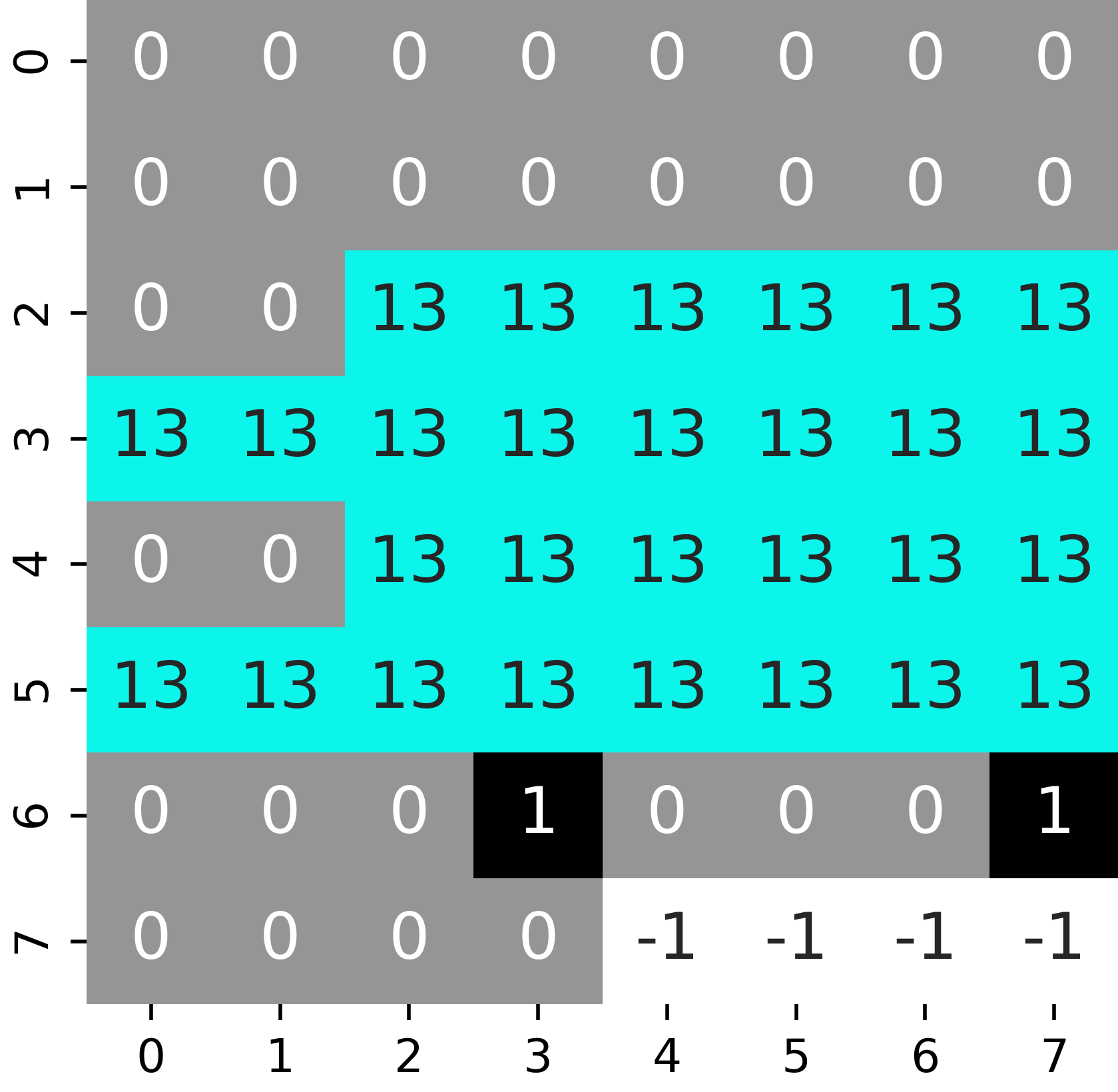}
  \end{minipage}
  \begin{minipage}[c]{0.62\textwidth}
  \caption{AID3 (identical to AID4): Bytes 2-3, 3-4 encode vehicle speed (DID13). 
  This is in fact wheelspeed for two wheels (AID4 encodes the other two). The last nibble is unmatched, but by visual inspection this is clearly a 4-bit Counter used to defend against injection and replay type attacks.
  Annotated open DBCs
(\url{https://community.comma.ai/cabana/})
indicate that it is common to have a 4-bit Counter ending a safety critical message.}
  \label{fig:mp-mid}
  \end{minipage}
\end{subfigure}

\hrule
\vspace{.2cm}

\begin{subfigure}[c]{0.48\textwidth}
    \centering
    \includegraphics[width =\textwidth]{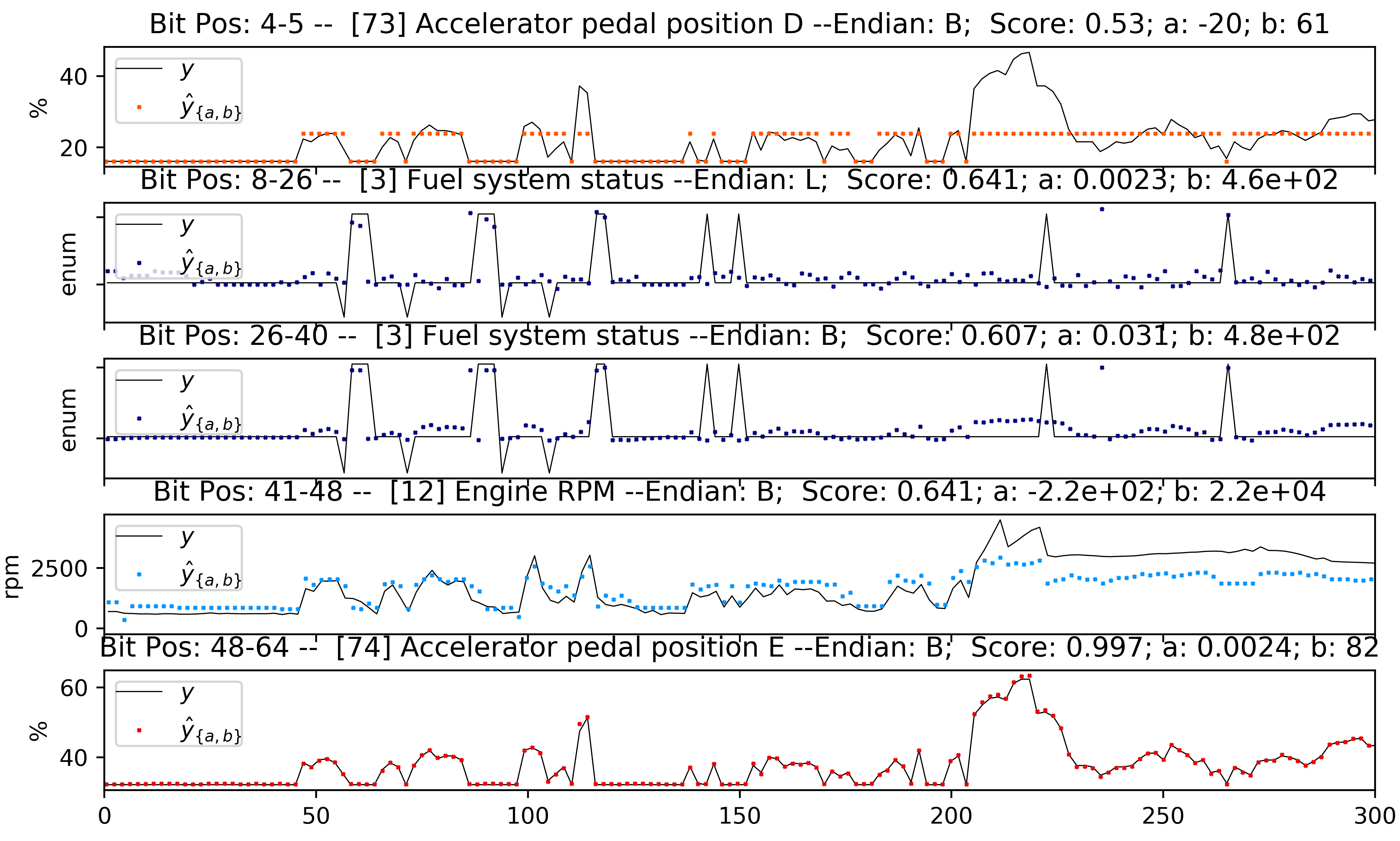}
  \begin{minipage}[c]{0.36\textwidth}
    \includegraphics[width =\textwidth]{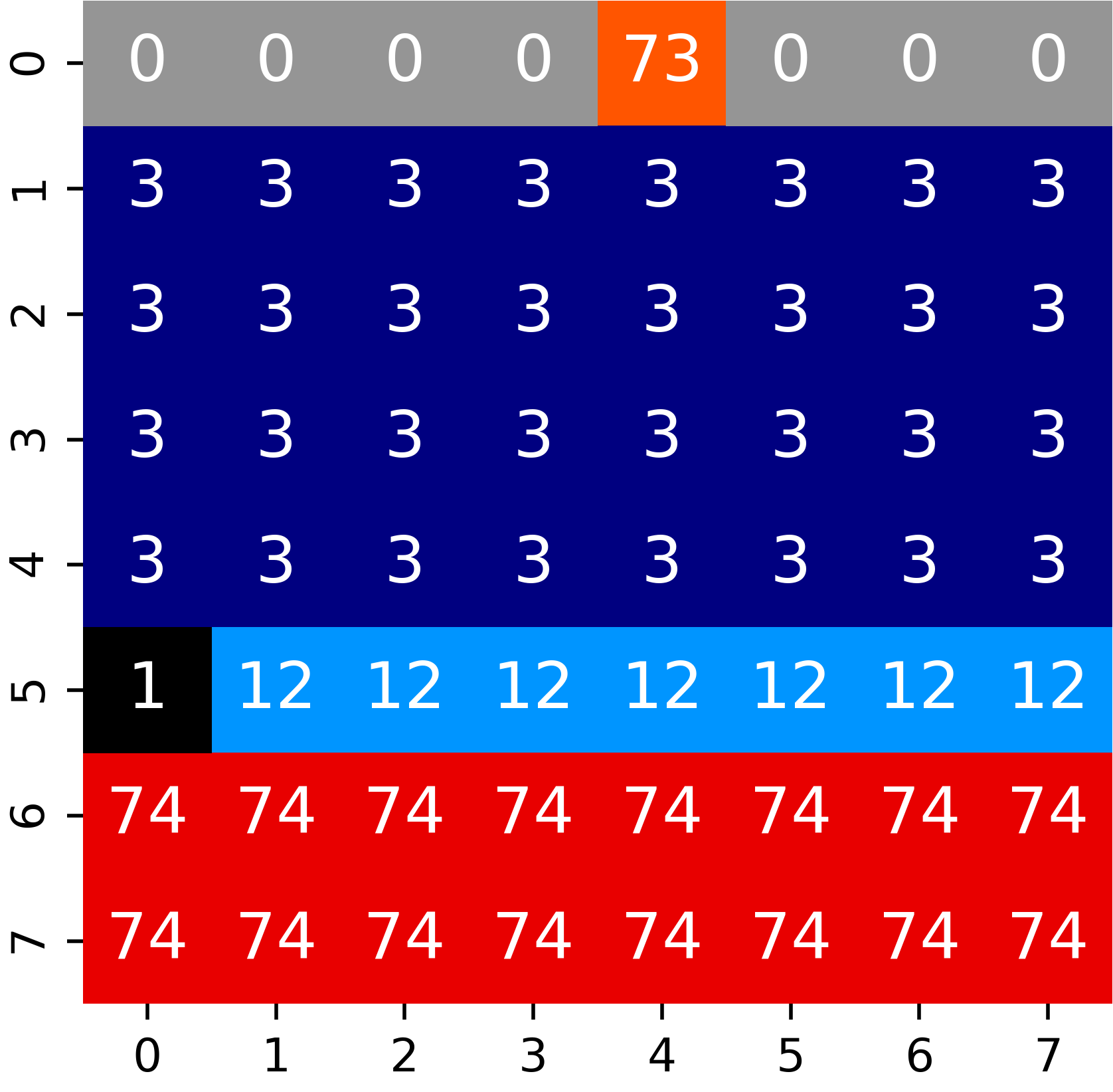}
    \end{minipage}
  \begin{minipage}[c]{0.62\textwidth}
  \caption{AID10: Matched tokens found in this message are (with the exception of final DID74 match) are signals that are likely correlated to the matched DID, but not an exact match. This is reflected in the significantly lower scores. Compare {\color{RedOrange} orange} and {\color{RoyalBlue} light blue} signals to those in \ref{fig:mp-top}. In \ref{fig:mp-bot} (top) the single bit mapped to DID73 seems to be a binary signal indicating whether the gas pedal is pressed. 
  }
  \label{fig:mp-bot}
  \end{minipage}
\end{subfigure}
\end{figure}

We look at three examples of mapped AID traces for this capture to illustrate results, shown in Figure \ref{fig:mappedpayloads}. 
Note that the AIDs are anonymized by replacement with their priority ranks (highest priority AID1, lowest is AID25). 
The DIDs are the actual OBD-II PIDs corresponding  to the documentation \cite{obd2pids}, and we have translated both the DIDs and mapped CAN tokens to the appropriate units. 
Note additionally that although all token boundaries in the shown examples are constant bits, this is not always the case. 

Figure \ref{fig:mp-top} presents a fully mapped message, that is, every bit is either constant or matched. 
The two matched tokens have very high match scores giving strong evidence that these two are exactly the signals reporting ``Engine RPM'' and ``Accelerator Pedal Position D'' DIDs, respectively. 
Note the single outlying orange point at  $\approx150$s,  an artifact of the differing sampling rates. 

Figure \ref{fig:mp-mid}  shows a very high priority message, AID3, which is nearly identical to AID4. 
Both contain two 12-bit tokens that are the signals for  ``Vehicle speed'' DID. 
Based on our experience working with CAN data, we can quickly identify these as encoding wheel speed\textemdash likely for all four wheels, two on each  AID3 and AID4. 
We note that the values of the linear coefficients, $a \approx1/100$, $b \approx 0$ shows these signals are simply the scaled speed encoded with higher precision in the CAN data than DID responses. 
Compare this to the coefficients in Figure \ref{fig:mp-top}, whose message scores and visual match are similarly high, but the coefficients indicate a more complex conversion. 
The two constant 1s bits in byte 6 of Figure \ref{fig:mp-mid}   are likely flag tokens (that do not flip states during the course of the capture) corresponding to the state of each wheel. 
The final half byte of the message is an unmatched token (shown with -1s). 
However, visual inspection of the token plot (bottom) show values vacillating between the 4-bit minimum ($2^0=0$) and maximum ($2^{3-1}=15$) over a regular period, revealing that it is a Counter Token. 
The characteristics of Counter Tokens, as well as Checksum Tokens, are described in \cite{Marchetti_Stabili_2018} and are apparently added to prevent particular types of attacks in safety-critical messages (e.g., injections or replay attacks). 
We note that many of our unmatched tokens may correspond to these Counter or Checksum Tokens. 
While our algorithm does currently not account for these, they are easy to identify, and we plan to identify them initially along with the constant bits in future iterations. 

The message in Fig. \ref{fig:mp-bot}  demonstrates the point that a lower score does not necessarily indicate poor performance, but simply a correlated signal. 
It also illustrates why it is reasonable for a single DID to match more than one AID token. 
For example, the first and third token in this message match DIDs that are matched with a higher score in \ref{fig:mp-top}. 
The single-bit token matched to DID73: ``Accelerator Pedal Position D'' has only a 0.53 match since only 53\% of total variance of the DID signal is explained by the one bit. 
However, it seems clear that this bit encodes whether or not the gas pedal is depressed\textemdash an indicator  unavailable via diagnostic query, but easily derived from pedal angle. 
Likewise, the third matched token is likely physically related to ``Engine RPM'' DID, but unlike the second token in \ref{fig:mp-top}, it is not an exact match. 
The second and third tokens matched (with opposite endianness) to ``Fuel System Status'' DID, which is an enumerated DID\textemdash values indicate a particular state, not a physical value (see \cite{obd2pids} for details). Significant deviation from the enumerated DID in portions of the plot may indicate a spurious match or that it communicates a different signal from a related system. 
Finally, the plots of the token values for this AID highlight the complexity of the translation problem, namely, similarity in the signal variation due to the highly interrelated physical system makes accurate translation/semantic labeling difficult. 

Overall, 
from this capture, we find 69.6\% of bits are constant 1s or 0s, and of the remaining bits, our algorithm matches 16.8\% leaving 13.6\% of bits unknown. 
By summing the message scores for each AID and dividing by the number of AIDs, we get a total match score of 14.5\% out of the 16.8\% matched bits for an overall match score of 86.0\%. 
We note that the score should not be interpreted as the performance of the algorithm as illustrated by Figure \ref{fig:mp-bot} where one can learn correlations or related signals and infer signals that are not DID-encoded (e.g., Figure \ref{fig:mp-bot}(top)).   
Changing $\alpha$ to .2, we obtain 22.0\% of the bits are matched and only 8.4\% unmatched with a total match score of 16.1\% over 22.0\% (72.9\%). Hence, we obtain some less direct matches but more insight into unknown but correlated tokens.






%% file: 40-conclusion.tex
\section{Conclusions} 
\label{sec:Conclusion}
Data transmitted over the in-vehicle CAN network is a veritable mine of information regarding vehicular functions. However, the key to decode CAN data, specifically the way to tokenize and translate messages, is completely proprietary, and varies per make, model, year and trim. Consequently, non-OEM augmentation of vehicles is greatly hindered because it operates blind to CAN message syntax and semantics.
We develop ACTT, the first algorithm to simultaneously tokenize and translate CAN data, learning message-to-function mappings by leveraging diagnostic information.

Our results show that ACTT both tokenizes and functionally translates  CAN data fields providing needed meaning. 
Specifically, many matched tokens reveal near-perfect DID encodings, while remaining matched tokens are correlated to their matched diagnostic responses providing potentially useful groupings and facilitate actual inference of the signal (if not a direct match) from inspection in some cases. 
We expect ACTT to provide a needed step in unlocking the DBC-encodings to enable a wide variety of after-market research including CAN security, performance tuning, and driver or vehicle state studies.

%% file: 90-acks.tex
\section*{Acknowledgements}
\label{sec:acks} 
Special thanks to Micheal Iannacone and anonymous reviewers. Research sponsored by the Laboratory Directed Research and Development Program of Oak Ridge National Laboratory, managed by UT-Battelle, LLC, for the U. S. Department of Energy (DOE) and by the DOE, Office of Science, Office of Workforce Development for Teachers and Scientists (WDTS) under the Scientific Undergraduate Laboratory Internship (SULI) program. 